\documentstyle[pmart,twoside,fleqn,epsf]{article} 

\newcommand{\Spg}{\mbox{{\bf S}}}
\newcommand{\be}{\begin{equation}}
\newcommand{\ee}{\end{equation}}
\newcommand{\ben}{\begin{eqnarray}}
\newcommand{\een}{\end{eqnarray}}
\newcommand{\ra}{\rangle}
\newcommand{\la}{\langle}

\newcommand{\kk}{{\bf k}}

\begin{document}

%
%
\title{DIAGONALIZATION OF A BOSONIC QUADRATIC FORM USING CCM:\\
       APPLICATION ON A SYSTEM WITH TWO INTERPENETRATING SQUARE LATTICE ANTIFERROMAGNETS}

\author{Sven E.~Kr\"uger, Dirk Schmalfu\ss, Johannes Richter}

\address{Institut f\"ur Theoretische Physik\\
         Universit\"at Magdeburg\\
         PF 4120, D-39016 Magdeburg, Germany}

\date{\today}  

\maketitle                   
%
%
\pacs{75.50.Ee,75.30.Ds,31.15.Dv,42.50.Ls}

\begin{abstract}
While the diagonalization of a quadratic bosonic form can always
be done using a Bogoliubov transformation, the practical implementation
for systems with a large number of different bosons is a tedious analytical
task. Here we use the coupled cluster method (CCM) to exactly
diagonalise such complicated quadratic forms. This yields
to a straightforward algorithm which can easily be implemented
using computer algebra even for a large number of different bosons. 
We apply this method on 
a Heisenberg system with two interpenetrating square lattice antiferromagnets, which
is a model for the quasi 2D antiferromagnet Ba$_2$Cu$_3$O$_4$Cl$_2$.
Using a four-magnon spin wave approximation we get a complicated Hamiltonian
with four different bosons, which is treated with CCM.
Results are presented for magnetic ground state correlations.
\end{abstract}

%
%
\section{Introduction -- The Model}
It is always possible to diagonalize quadratic bosonic forms
(which appear frequently in physics) using a Bogoliubov transformation \cite{bogoliubov47}, but it
is a tedious analytical task to find one for a complicated form
with many different magnons. Therefore we want to show here how
the coupled cluster method (CCM), one of
the most powerful and universal techniques in quantum many-body theory
(s. \cite{bishop91} and references therein),
can be used in a straightforward scheme
to find the exact ground state of such a form.

To be concrete we consider the 2D spin $1/2$ Heisenberg model
\begin{equation} \label{ham}
  H=J_{AA}\sum_{\langle i\in A_1,j\in A_2\rangle}{\bf S}_{i}{\bf S}_{j}+
  J_{BB}\sum_{\langle i\in B_1,j\in B_2\rangle}{\bf S}_{i}{\bf S}_{j}+
  J_{AB}\sum_{\langle i\in A,j\in B\rangle}{\bf S}_{i}{\bf S}_{j} ,
\end{equation}
which is related to the situation in Ba$_2$Cu$_3$O$_4$Cl$_2$ \cite{richter98,richter99},
a layered quantum antiferromagnet showing
significant differences to its parent cuprats like La$_2$CuO$_4$
(see e.g.~\cite{chou97} for recent experiments).  In contrast
to La$_2$CuO$_4$ we have two different types of Cu-sites in the
Cu-O-planes, namely there are additional Cu(B) atoms located in
the centre of every second Cu(A)-O$_2$ square.
Within the Cu(A) subsystem we have a strong 180$^{\circ}$ Cu-O-Cu
superexchange yielding to strong antiferromagnetic couplings ($J_{AA}$) between
Cu(A) atoms, whereas the couplings within the Cu(B) subsystem
($J_{BB}$) and between the subsystems ($J_{AB}$) are weaker.
A recent calculation of $J_{AA}$, $J_{BB}$ \cite{rosner98}, finding
$J_{AA}\approx 10J_{BB}$ (both antiferromagnetic)
agrees with the experimental values \cite{chou97}.
There are also some arguments \cite{rosner98} for a ferromagnetic
$|J_{AB}|\approx J_{BB}$.

In the classical ground state (\ref{ham}) shows
for $|J_{AB}|\leq 2\sqrt{J_{AA}J_{BB}}$
a N\'eel like order for the two
subsystems A and B, where the energy is degenerated with respect
to the angle $\varphi$ between the spins of these two
subsystems.

\section{The Method}
In this paper we study the ground state properties of (\ref{ham}),
using a four-magnon linear spin wave approximation \cite{holstein40}
around the classical ground state, i.e.~for
each of the four sublattices $A_1,A_2,B_1,B_2$
of the two coupled bipartite antiferromagnetic
square lattices we introduce different bosonic operators.
Thus we get for (\ref{ham})
\be \label{ham_lswt}
H=-\frac {2N}{3}s^2\left(2J_{AA}+J_{BB}\right)+\sum_{{\bf k}}H_{{\bf k}}, \quad\mbox{with}
\ee
\be \label{hk}\begin{array}{l}
H_{{\bf k}}=
  4J_{AA}s\left(a_{1{\bf k}}^{+}a_{1{\bf k}}+a_{2{\bf
k}}^{+}a_{2{\bf k}}-\gamma_{{\bf k}AA}\left[a_{1{\bf k}}^{+}a_{2-{\bf
k}}^{+}+a_{1{\bf k}}a_{2-{\bf
k}}\right]\right) \\
 +2J_{BB}s\left(b_{1{\bf k}}^{+}b_{1{\bf k}}+b_{2{\bf
k}}^{+}b_{2{\bf k}}-\gamma_{{\bf k}BB}\left[b_{1{\bf k}}^{+}b_{2-{\bf
k}}^{+}+b_{1{\bf k}}b_{2-{\bf
k}}\right]\right) \\
 +J_{AB}s(1+\cos\varphi)/2\left(b_{1{\bf
k}}^{+}a_{1{\bf k}}+b_{1{\bf
k}}a_{1{\bf k}}^{+}-b_{2{\bf k}}^{+}a_{1-{\bf k}}^{+}
-b_{2{\bf k}}a_{1-{\bf k}}\right)\gamma_{{\bf k}AB}^{1} \\
 +J_{AB}s\left(1-\cos\varphi\right)/2\left(b_{2{\bf
k}}^{+}a_{2{\bf k}}+b_{2{\bf
k}}a_{2{\bf k}}^{+}-b_{1{\bf k}}^{+}a_{2-{\bf k}}^{+}
-b_{1{\bf k}}a_{2-{\bf k}}\right)\gamma_{{\bf k}AB}^{2} \\
 +J_{AB}s\left(1-\cos\varphi\right)/2\left(b_{2{\bf
k}}^{+}a_{1{\bf k}}+b_{2{\bf
k}}a_{1{\bf k}}^{+}-b_{1{\bf k}}^{+}a_{1-{\bf k}}^{+}
-b_{1{\bf k}}a_{1-{\bf k}}\right)\gamma_{{\bf k}AB}^{1} \\
 +J_{AB}s\left(1+\cos\varphi\right)/2\left(b_{1{\bf
k}}^{+}a_{2{\bf k}}+b_{1{\bf
k}}a_{2{\bf k}}^{+}-b_{2{\bf k}}^{+}a_{2-{\bf k}}^{+}
-b_{2{\bf k}}a_{2-{\bf k}}\right)\gamma_{{\bf k}AB}^{2},
\end{array} \ee
using the lattice structure factors $\gamma_{{\bf k}AA}=\cos(k_x/2)\cos(k_y/2)$,
$\gamma_{{\bf k}BB}=(\cos k_x+\cos k_y)/2$ and
$\gamma_{{\bf k}AB}^{1(2)}=\cos(k_{x(y)}/2)$.

As stated, we use the coupled cluster method (CCM)
to find the exact ground state of (\ref{ham_lswt}).
To do this we notice the following property of $H$
\be
  \sum_{\kk}H_{\kk}=\sum_{\kk}(H_{\kk}+H_{-\kk})/2\equiv\sum_{\kk}H_{\kk}'; \qquad \Rightarrow [H_{\kk}',H_{\kk'}']_-=0 \quad \forall \kk,\kk' .\ee
Hence it is possible to treat
each $H_{\kk}'$ seperately within the CCM, since they all commute with each other.
So we have to deal with a bosonic system with eight different bosonic operators
$a_{1\pm\kk}$, $a_{2\pm\kk}$, $b_{1\pm\kk}$, $b_{2\pm\kk}$ denoted
with $a_1,\dots,a_8$.

The ket and bra ground state of such a system
(i.e~ a many-mode bosonic field theory with bosonic operators
$a_i$, $a_i^+$ in the Hamiltonian) in CCM-SUB$l$ approximation
is given by \cite{bishop91,arponen91}
\be\label{ccm_state} \begin{array}{ll}
  |\Psi\ra =e^S|0\ra, & S=\sum_{i_1,i_2,\dots,i_l}A_{i_1,i_2,\dots,i_l}a_{i_1}^+a_{i_2}^+\cdots a_{i_l}^+, \\
  \la\tilde\Psi|=\la 0|\tilde Se^{-S}, & \tilde S=1+\sum_{i_1,i_2,\dots,i_l}\tilde A_{i_1,i_2,\dots,i_l}a_{i_1}a_{i_2}\cdots a_{i_l}
\end{array}  ,\ee
where $|0\ra$ is the bosonic vacuum state (i.e.~$a_i|0\ra=0$), and
$A_{i_1\cdots}$ and $\tilde A_{i_1\cdots}$ are the CCM correlation coefficients.
These coefficients are calculated by two systems of equations (one
of them is a system of nonlinear equations).
\be\label{ccm_eq}
  \frac{\partial \bar H}{\partial \tilde A_{i_1\cdots i_l}}=0, \quad \frac{\partial \bar H}{\partial A_{i_1\cdots i_l}}=0
   ,\quad \bar H=\la\tilde \Psi|H|\Psi\ra ,\ee
using the expectation value ($\bar H$) of the Hamiltonian, i.e.~the ground state energy.

Note, that the CCM-SUB2 approximation
(i.e.~having only quadratic terms
of bosonic operators in $S$ and $\tilde S$ (\ref{ccm_state}))
gives the {\em exact} ground state of a {\em quadratic} bosonic Hamiltonian, since
the ground state wave function of such a Hamiltonian has the form
$|\Psi\ra=\exp[\sum_{ij}f_{ij}a_i^+a_j^+]|0\ra$, which can easily be shown
using a Bogoliubov transformation (see appendix). Therefore the CCM correlation operator
$S$ (and $\tilde S$ respectively) (\ref{ccm_state}) consist of products
of {\em two} bosonic creation operators only, all other coefficients
$A_{i_1,\cdots,i_l}$ are zero; so we just have to use SUB2.

To calculate the CCM equations (\ref{ccm_eq}) easily using computer algebra, we make use
of the Bargmann representation \cite{bargmann61,arponen91}
\be\label{barg}
  a^+\Leftrightarrow z,\quad a\Leftrightarrow \frac{d}{dz}, \quad |0\ra\Leftrightarrow 1,
  \quad \la 0|f(a,a^+)|0\ra\Leftrightarrow \left. f(\frac{d}{dz},z)\right|_{z=0}
,\ee
which maps the original many-mode bosonic field theory into the
corresponding (classical) field theory of complex functions in
a particular normed space. So instead of bosonic operators we just
have to handle with (complex) numbers and differential operators, which is
much easier. Once the (partial nonlinear) equations are obtained they can be
solved numerically.

\section{Results and conclusions}
\begin{figure}[t]
  \centerline{
    \epsfysize=4.5cm \hspace{3cm}
    \epsfbox{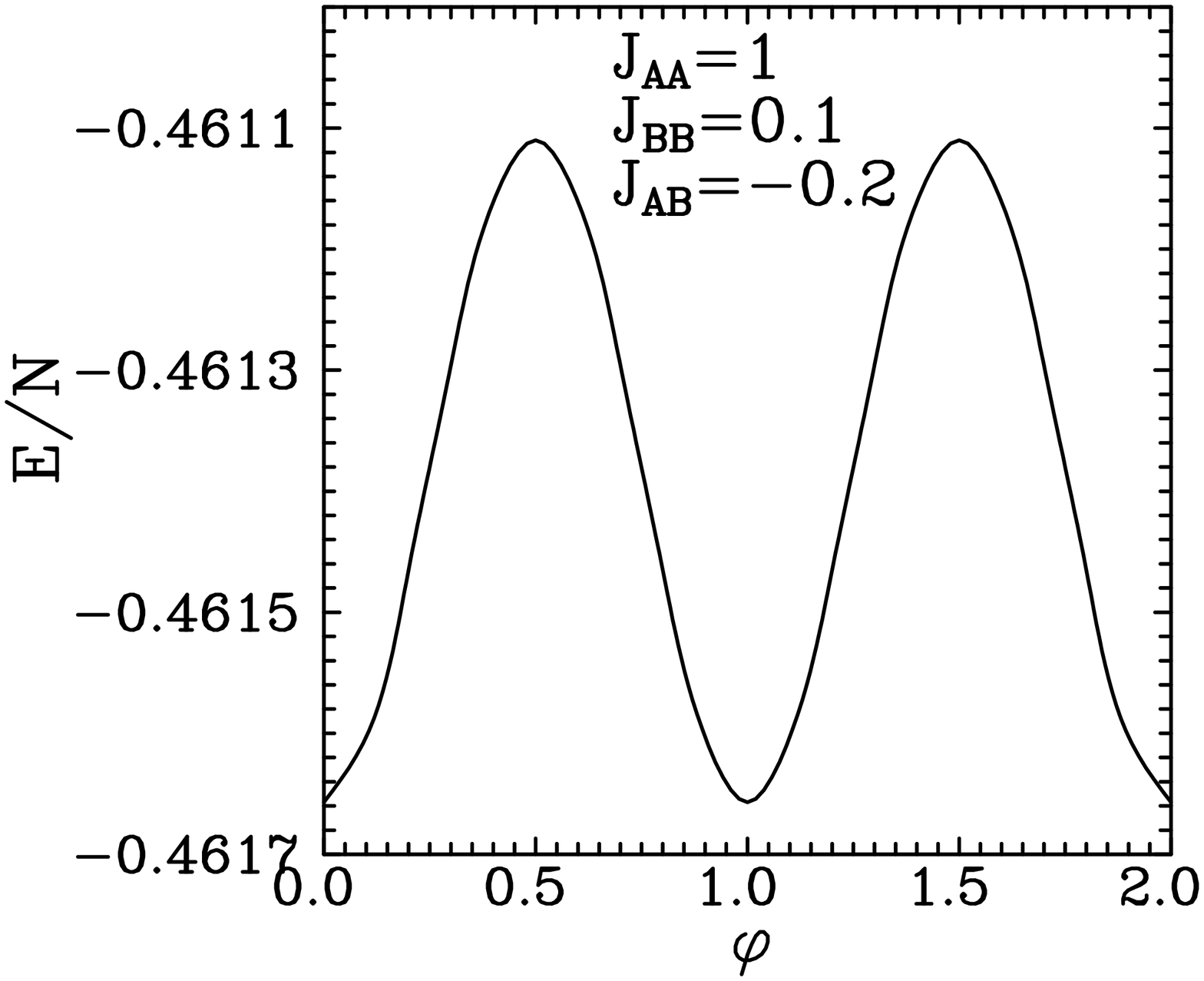}
    \epsfysize=4.5cm  \hspace{-3cm}       
    \centerline{\epsfbox{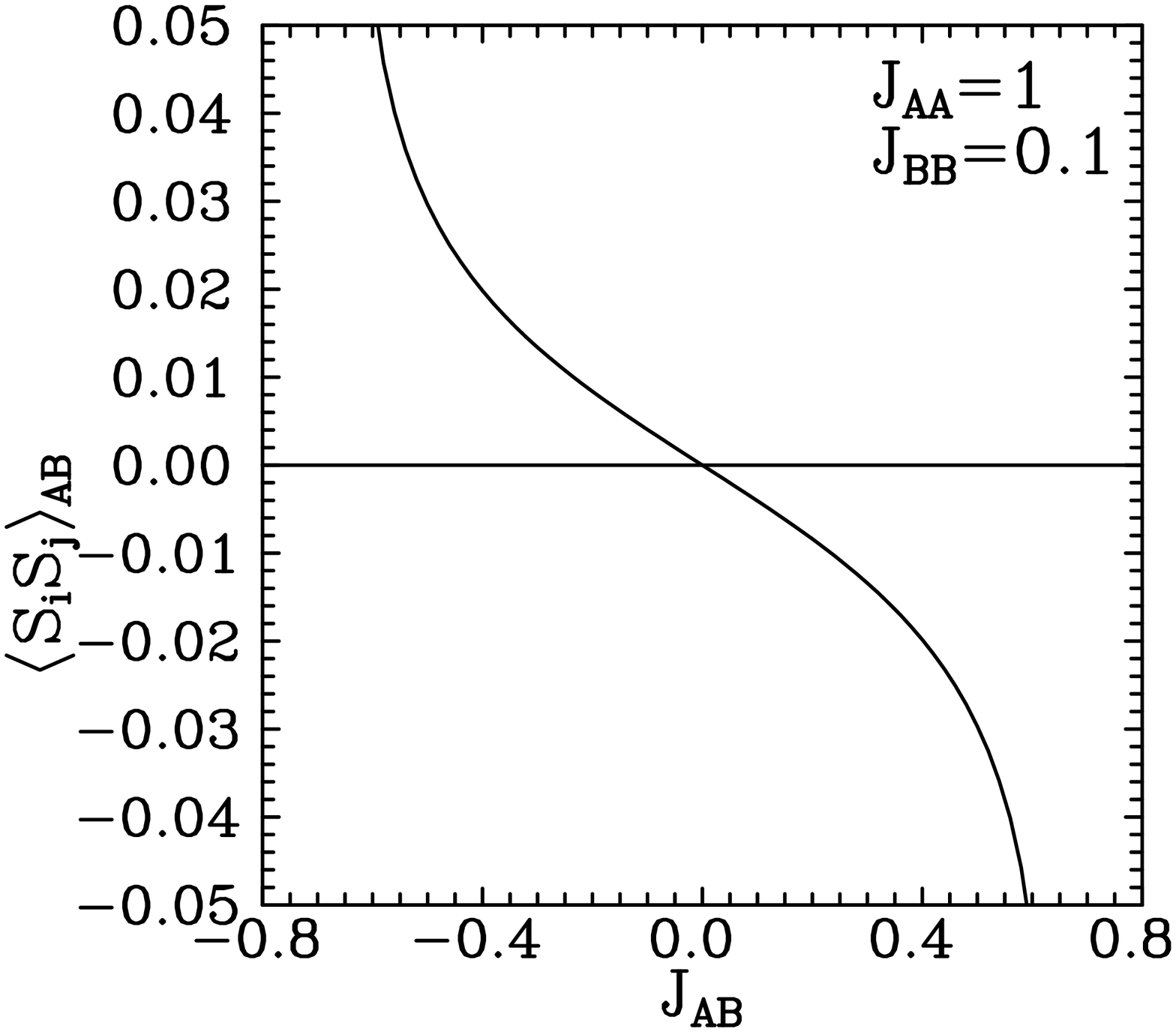}}
  }    
  \caption{Ground state of the model (\protect{\ref{ham}}): 
           (a) energy as a function of
           the (classical) angle $\varphi$ (in units of $\pi$)
           between the two subsystems A and B
           (note, that in the classical case the energy does not depend on $\varphi$);
           and (b) correlation between spins of these 
           two subsystems in dependence on $J_{AB}$.}
  \label{fig}
\end{figure}

We apply the CCM-scheme described above to calculate the exact ground state
of (\ref{hk}) and by doing this getting a spin wave
approximate ground state of the model (\ref{ham}).
We discuss the energy as a function of the
angle between spins of the two subsystems $A$ and $B$
and the correlation
between spins of different subsystems as a function of $J_{AB}$
(Fig.\ref{fig}). We find
as a typical {\em order from disorder} effect, that the degeneracy of the
ground state with respect to the angle $\varphi$ is lifted by quantum
fluctuations and a collinear ordering ($\varphi=0,\pi$) is stabilized.
This can clearly be seen by the energy vs.~$\varphi$ picture in Fig.\ref{fig}
and by the correlation $\la\Spg_i\Spg_j\ra_{A,B}$ vs.~$J_{AB}$, which
is zero in the classical case, independent of the value of $J_{AB}$
(for $|J_{AB}|\leq 2\sqrt{J_{AA}J_{BB}}$).
In the quantum case however that correlation does depend on $J_{AB}$,
showing again an order effect induced by quantum fluctuations.

In addition we find a lowering of the magnetic order within the subsystems
$A$ and particular $B$ by frustrating $J_{AB}$ in the quantum case.

\section*{Acknowledments}
This work has been supported by the DFG (Project Nr. Ri 615/7-1).

\begin{appendix}
\section{Proof that CCM-SUB2 gives exact ground state}
Using the fact, that a Bogoliubov transformation
$\beta_{\nu}=\sum_{\mu}(u_{\mu\nu}^*a_{\mu}-v_{\mu\nu}^*a_{\mu}^+)$
exactly diagonalize a quadratic bosonic Hamiltonian with the bosonic operators
$a_i$, $a_i^+$, one can easily show that its ground state must have the form
$|\Psi\ra=\exp[\sum_{ij}f_{ij}a_i^+a_j^+]|0\ra$,
by showing that $\beta_{\nu}|\Psi\ra=0$ $\forall \nu$.
We use the Bargmann representation (\ref{barg}) and get
\[ \beta_{\nu}|\Psi\ra\stackrel{!}{=}0
          \Leftrightarrow \sum_{\mu}\left(u_{\mu\nu}^*\frac{d}{dz_{\mu}}-v_{\mu\nu}^*z_{\mu}\right)\exp\left[\sum_{ij}f_{ij}z_iz_j\right]\stackrel{!}{=}0 \quad \forall z_i \]
\[ \Rightarrow \sum_{\mu}(u_{\mu\nu}^*2\sum_if_{i\mu}z_i-v_{\mu\nu}^*z_{\mu})\stackrel{!}{=}0 \quad \forall z_i, \quad \Rightarrow 2\sum_{\mu}f_{i\mu}u_{\mu\nu}^*\stackrel{!}{=}v_{i\nu}^* \]
and this last matrix equation is allways fulfilled for some $f_{i\mu}$.

\end{appendix}


\begin{thebibliography}{}
  
\bibitem{chou97} F.C. Chou, A. Aharony, R.J. Birgeneau, O. Entin-Wohlman,
 M. Greven, A.B. Harris, M.A. Kastner, Y.J. Kim, D.S. Kleinberg, Y.S. Lee
 and Q. Zhu, {\em Phys. Rev. Lett.} {\bf 78}, 535, (1997)  
\bibitem{rosner98} H. Rosner,
   {\em Phys. Rev. B} {\bf 57}, 13660 (1998)  
\bibitem{richter98} J. Richter, N.B. Ivanov, R. Hayn, J. Schulenburg,
   {\em J. Magn. Magn. Mat.} {\bf 177-181}, 737 (1998)  
\bibitem{richter99} J. Richter, D. Schmalfu\ss, S. Kr\"uger,
   {\em Physica B} {\bf 259-261}, 911 (1999)

\bibitem{bishop91} R.F. Bishop,
        {\em Theor. Chim. Acta} {\bf 80}, 95 (1991).
\bibitem{arponen91} J.S. Arponen, R.F. Bishop,
        {\em Theor. Chim. Acta} {\bf 80}, 289 (1991).  

\bibitem{bargmann61} V. Bargmann, {\em Comm. Pure Appl. Math.} {\bf 14}, 180, 187 (1961);
                     ibid idem {\bf 20}, 1 (1967);
                     V. Bargmann, P. Butera, L. Girardello, J.R. Klauder,
                          {\em Rep. Math. Phys.} {\bf 2}, 221 (1971)

\bibitem{bogoliubov47} N. Bogoliubov, {\em J. Phys. (USSR)} {\bf 11}, 23 (1947);
                       J.H.P. Colpa, {\em Physica} {\bf 93A}, 327 (1978)

\bibitem{holstein40}
   T. Holstein, H. Primakoff, {\em Phys. Rev.} {\bf 58}, 1908 (1940)

\end{thebibliography}
\end{document}